\documentstyle[11pt,epsf,aaspp4]{article}
\def\Rin{R_{\rm in}}
\def\Rg{R_{\rm g}}

\def\mdcrit{{\dot m}_{\rm crit}}
\def\msun{M_{\odot}}
\def\dm{{\dot m}}
\def\ref{{\hang\noindent}}

\def\plotone#1{\centering \leavevmode
\epsfxsize=0.45\textwidth \epsfbox{#1}}
\def\plottwo#1#2{{\centering \leavevmode
\epsfxsize=0.45\textwidth \epsfbox{#1}} \centering \leavevmode
\epsfxsize=0.45\textwidth \epsfbox{#2}}

\righthead{}
\lefthead{}

\begin{document}

\title{The relativistic iron line profile in the Seyfert 1 Galaxy IC4329a}

\author{C. Done$^1$, G.M. Madejski$^2$, P.T. \.{Z}ycki$^3$}
\affil{$^1$ Department of Physics, University of Durham, South Road,
	Durham DH1 3LE, England; chris.done@durham.ac.uk}
\affil{$^2$ NASA/Goddard Space Flight Center, Greenbelt, MD 20771, USA; madejski@lheavx.gsfc.nasa.gov}
\affil{ $^3$ Nicolaus Copernicus Astronomical Center, Bartycka 18,
            00-716 Warsaw, Poland; ptz@camk.edu.pl}

\begin{abstract}

We present simultaneous ASCA and RXTE data on the bright Seyfert 1
galaxy IC4329a. The iron line is significantly broadened, but not to
the extent expected from an accretion disk which extends down to the
last stable orbit around a black hole. We marginally detect a narrow
line component, presumably from the molecular torus, but even
including this gives a line profile from the accretion disk which is
significantly narrower that that seen in MCG--6--30--15, and is much
more like that seen from the low/hard state galactic black hole
candidates.  This is consistent with the inner disk being
truncated before the last stable orbit, forming a hot flow at
small radii as in the ADAF models. However, we cannot rule out the
presence of an inner disk which does not contribute to the reflected
spectrum, either because of extreme ionisation suppressing the
characteristic atomic features of the reflected spectrum or because
the X--ray source is intrinsically anisotropic, so it does not
illuminate the inner disk.

The source was monitored by RXTE every 2 days for 2 months, and these
snapshot spectra show that there is intrinsic spectral variability.
The data are good enough to disentangle the power law from the
reflected continuum and we see that the power law softens as the
source brightens.  The lack of a corresponding increase in the
observed reflected spectrum implies that either the changes in disk
inner radial extent/ionisation structure are small, or that the
variability is actually driven by changes in the seed photons which
are decoupled from the hard X--ray mechanism.

\end{abstract}


\section{INTRODUCTION}

Accretion of material onto a black hole is known to produce hard
X--ray ($E \ge 2$ keV) emission: satellite observations over the last 30 
years have
conclusively shown this to be the case for both stellar mass black
holes in our own Galaxy, and the supermassive black holes which power
the Active Galactic Nuclei (AGN).  However, it is still the case that
the mechanism by which the gravitational potential energy is converted
into high energy radiation is not understood.  Standard models of an
accretion disk (Shakura \& Sunyaev 1973, hereafter SS) produce copious
UV and even soft X--ray radiation, but are completely unable to
explain the observed higher energy X--ray emission which extends from 
the disk spectrum out to 200 keV. Clearly something other
than the standard model is required, and the two currently favored
candidates are either magnetic reconnection above an accretion disk
(e.g. Galeev, Rosner \& Vaiana 1979; Haardt, Maraschi \& Ghisellini
1994; di Matteo 1998), or that some part of the accretion flow is not
given by the standard disk configuration.

The recent rediscovery of another stable solution of the accretion
flow equations lent plausibility to this second possibility. The
standard SS accretion disk model derives the accretion flow structure
in the limit where the gravitational energy released is radiated
locally in an optically thick, geometrically thin disk. This contrasts
with the new accretion models, where below some critical mass
accretion rate, $\dm\le \mdcrit$, the material is not dense enough to
thermalise and locally radiate all the gravitational potential energy
which is released. Instead the energy can be carried along with the
flow (advected), eventually disappearing into the black hole. These
solutions give an X--ray hot, optically thin, quasi--spherical flow
(Narayan \& Yi 1995; Esin, McClintock \& Narayan 1997).

Clearly it would be nice to know which (if any!) of these models for
the hard X--ray emission is correct. In its most general form the
problem comes down to understanding the geometry of the cool,
optically thick accretion flow. If this extends down to the last
stable orbit around the black hole then it is unlikely that the
advective flow can exist (e.g.\ Janiuk, \.{Z}ycki \& Czerny 1999). 
Conversely, if the optically thick disk
truncates before this point then a composite model with an outer disk
and inner, hot, advective flow (ADAF) may be favored. 

The accretion flow can be tracked via X--ray spectroscopy. Wherever hard
X--rays illuminate optically thick material then this gives rise to a Compton
reflected continuum and associated iron K$\alpha$ fluorescence line (e.g.\
Lightman \& White 1988; George \& Fabian 1991; Matt, Perola \& Piro 1991).  The
amplitude of the line and reflected continuum is dependent on the solid angle
subtended by the disk to the X--ray source (as well as on elemental 
abundances, inclination and ionisation). The two models outlined above 
can then be distinguished by the amount of reflection and line, since 
an untruncated disk should subtend a rather larger solid angle 
than a truncated one. 

A survey of AGN showed that Seyfert 1 spectra are consistent with a
power law X--ray spectrum illuminating an optically thick, (nearly)
neutral disk, which subtends a solid angle of $\sim 2\pi$ (Pounds et
al. 1990; Nandra \& Pounds 1994). This then seems to favor the
magnetic reconnection picture. However, this contrasts with the
situation in the Galactic Black Hole Candidates (GBHC). These are also
thought to be powered by accretion via a disk onto a black hole, and,
in their low/hard state, show broad band spectra which are rather
similar to those from AGN, but have an apparently rather smaller
amount of reflection (e.g. Gierli\'{n}ski et al. 1997; Done \&
\.{Z}ycki 1999). One potential drawback of this comparison
is that AGN inhabit a more complex environment than the GBHC.
Unification schemes for Seyfert galaxies propose that there is a
molecular torus which enshrouds the nucleus. The molecular torus can
also contribute to the reflected spectrum, perhaps distorting our view
of the very innermost regions in AGN.

These two potential sites for the reflected component can be distinguished
spectrally: any features from the accretion disk should be strongly smeared by
the combination of special and general relativistic effects expected from the
high orbital velocities in the vicinity of a black hole (Fabian et al.  1989),
whereas the molecular torus is at much larger distances so its reflected
features should be {\it narrow}. The seminal ASCA observation of the AGN
MCG--6--30--15 showed that the line is so broad that it requires that the
accretion disk extends down to {\it at least} the last stable
orbit in a Schwarzschild metric, with {\it no} narrow component from the
molecular torus (Tanaka et al. 1995; Iwasawa et al. 1996), and that the
relativistically smeared material subtended a solid angle of $\sim 2\pi$ with
respect to the X--ray source.  This very clear cut result then seems
to rule out the advective flows, at least in their simplest form.  

However, the GBHC again show a rather different picture: they have a
line which is detectably broad, but not so broad as might be expected
for a disk extending all the way down to 3 Schwarzschild radii 
($R_{\rm Schw} = 2 G M/c^2$; 
\.{Z}ycki, Done, \& Smith 1997, 1998, 1999; Done \& \.{Z}ycki 1999).  This
{\it is} consistent with the truncated disk geometry, and so perhaps
with the advective flow models.  Only in the soft/high state do the
GBHC seem to show the extreme relativistic smearing and large amount
of reflection seen in the MCG--6--30--15 spectra (\.{Z}ycki et al. 1998,
Gierli\'{n}ski et al. 1999).

Is there a real difference in accretion geometry between GBHC and AGN,
pointing to a difference in radiation mechanism ? This seems unlikely,
since both classes are ultimately accreting black holes. Are subtle
ionisation effects masking the derived disk parameters in the GBHC
(Ross, Fabian \& Young 1999).  Or is MCG--6--30--15 unusual among AGN
(and GBHC) in having such a relativistic disk?  Perhaps MCG--6--30--15
is an AGN in a state which corresponds to the soft state GBHC?

One factor supporting the latter is a recent study by Zdziarski, 
Lubi\'{n}ski, \& Smith (1999) which showed that there is a 
correlation between the intrinsic spectral slope, $\Gamma$, and 
the solid angle subtended by the reflecting material, 
$\Omega/2\pi$. In their plots, MCG--6--30--15 is the 
AGN with the steepest spectrum, and highest amount of reflection.  
This correlation also holds for individual objects (such as NGC 5548 
for AGN and Nova Muscae for the GBHC), where the intrinsic 
spectrum hardens as the amount of reflection decreases.  This 
suggests that there is a universal physical mechanism and/or 
geometry for both classes, with perhaps a single parameter 
determining the state of a given source through a feedback between the
geometry and physical conditions in the X--ray emitting region.  
Such a feedback could be provided by e.g.\ soft photons from 
the thermalized fraction of the hard X--rays intercepted by the 
reprocessing medium, and the parameter could be the inner disk 
radius.  Perhaps for MCG--6--30--15 (and other soft AGN and the
soft state GBHC) the cool accretion disk extends down to the innermost 
stable orbit around the black hole with the X--rays being powered 
by magnetic reconnection above the disk, while for harder spectra 
AGN (and the low state GBHC) the inner disk recedes outwards, 
being replaced by an X--ray hot (advective ?) flow.  As
the disk recedes it subtends a smaller solid angle, so 
there is less reflection (and less relativistic smearing), but 
there are also fewer seed photons from the disk (both from
intrinsic emission and by reprocessing) for Compton scattering 
into the intrinsic power law, giving a harder intrinsic power 
law (\.{Z}ycki et al. 1999;  Zdziarski et al. 1999).

If this is true, then this clearly predicts that the Fe K$\alpha$ 
line {\it in AGN} is not always as broad as in the extreme case of 
MCG-6-30-15 (Tanaka et al.\ 1995).  Previous ASCA studies 
on a sample of objects (Nandra et al.\ 1997) hint towards such a 
possibility, since a whole range of geometries were inferred for
various objects. 
 
Here we test this idea using ASCA, XTE and OSSE data from IC 4329a, 
the brightest 'typical' Seyfert 1 in the X--ray band (cf. Madejski et
al. 1995). It lies towards the middle of the $\Gamma-\Omega/2\pi$ plot 
of Zdziarski et al. (1999), and (consequently) has a spectrum very 
close to that of the mean Seyfert 1 spectrum compiled by Zdziarski 
et al. (1995). IC 4329a may then be used as a template for
Seyfert galaxies as a class, unlike MCG--6--30--15 which has a rather steep
X--ray spectral index. 

\section{DATA REDUCTION}

\subsection{ASCA}

The 1997 campaign for IC 4329a included
four ASCA pointings, on August 7, 10, 12, and 16, each
nominally providing 20 ks of data.  The resulting data were extracted
using the standard screening procedures, yielding total exposures of
61 ks for SIS0 and SIS1 (using the {\tt BRIGHT2} mode), and 78 ks for 
GIS2 and GIS3.  The source data were
extracted from circular regions with radii of 3 arc min for the SISs
and 4 arc min for the GISs, while background was taken from a
source-free regions of the same images.  
The source showed a clear variability between the
four pointings, with GIS2 counting rates of $1.77 \pm 0.009$, 
$1.27 \pm 0.008$, $1.53 \pm 0.009$, and $1.78 \pm 0.009$, matching the 
variability seen in the simultaneous RXTE data (see below and Fig. 1).   
The PHA data were then grouped so as to have more than 20 counts per 
bin.  

\subsection{RXTE: PCA}

IC4329a was observed a total of 66 times over a period of 58 days with
RXTE. The PCA standard 2 data were extracted from all layers of
detectors 0,1 and 2, using standard selection criteria (Earth
elevation angle $> 10^\circ$, offset between the source and the
satellite pointing direction $< 0.02^\circ$, electron rates in each
detector $< 0.1$, excluding data taken within 30 minutes of the SAA
passage). This gave a total of 73 ks of good data. The background for
these data were then modeled using the 'L7' model (see Zhang et
al. 1993;  Madejski et al. 1999;  and Jahoda et al. 2000, in preparation), and
the corresponding response matrix for each observation was generated
using version 3.0 of the channel to energy conversion file.  We use
data from 3--20 keV, since lower energies are affected by response
matrix uncertainties and the background may not be well modeled above
20 keV. A 1\% systematic error is applied to data in all PHA
channels.  

\subsection{RXTE: HEXTE}

The HEXTE instrument onboard RXTE (cf. Rothschild et al. 1998)
consists of two scintillator modules, sensitive in the hard X--ray
band.  The background is measured via 
chopping the detectors to off-axis, source free locations every 
16 seconds.  The HEXTE data were extracted using standard data 
reduction procedures, which, after appropriate dead-time correction, 
yielded a total net exposure of $\sim 24$ ks.  The source was 
clearly detected in each pointing over the range of 15 -- 100 keV
range, at a level consistent with 
the PCA, but the statistical errors in each pointing were too
large to study variability.  The resultant data were binned such that 
channels 17--28 (16.9-28.3 keV) 
were grouped by a factor 3, 29--48 (28.3--47.8 keV) by a factor 4 and
49--120 (47.8--123.4 keV)
by a factor 6.  The effective area of both HEXTE clusters was
scaled by 0.7, the current best normalization to the Crab spectrum.  

\subsection{OSSE}

OSSE pointed at IC~4329a during the CGRO viewing period 625, over the
epoch of 1997 August 8 to 18, with the total exposure of 567 ks.  
The data were reduced in a standard manner (see Johnson et al. 1997 and
references therein), resulting in a net counting rate of $0.18 \pm
0.08$ counts s$^{-1}$ over the 50 -- 500 keV range.  The resulting 
data were binned such that channels 9--18 were grouped by a factor 
5, while 19--48 were grouped by a factor 10.  

\subsection{Lightcurves}

Figure 1 shows the 2--10 keV light curve obtained from the RXTE PCA
data. The source is clearly variable on timescales of a few days, with 
an {\it r.m.s} variation of 13\%. There is no significant 
short timescale variability within each observations. These typically
have 3$\sigma$ upper limits of 0.06 to the fractional {\it r.m.s}
variability of the 1000--2000 second lightcurve binned on 16 seconds.
The horizontal lines on Figure 1 mark the times at which ASCA and OSSE 
data were taken.

\begin{figure*}
\plotone{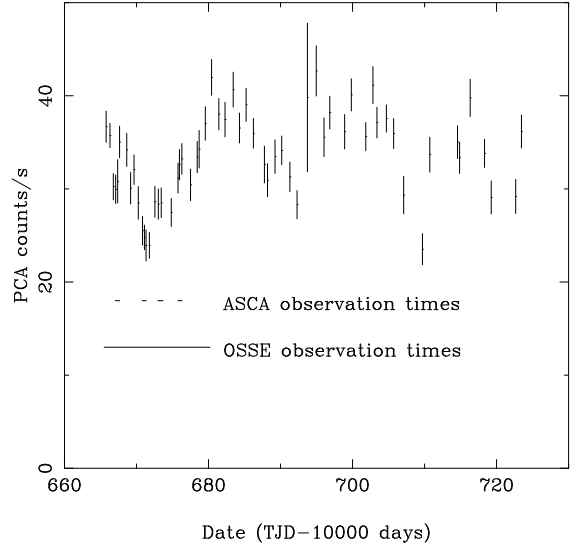}
\figcaption{XTE PCA count rate showing the variability throughout the 
monitoring campaign. The times of the ASCA and OSSE observations are 
marked by the horizontal lines}
\end{figure*}

\section{SPECTRAL FITTING}

The data were fitted using XSPEC v10.0, with errors quoted as 90\%
confidence intervals ($\Delta\chi^2=2.7$).  We use elemental 
abundances and cross-sections of Morrison \& McCammon (1983).

\subsection{ASCA SIS and GIS}

Data from SIS0 and SIS1 
are showing increasing divergence with time from the GIS2 and GIS3 (and
from each other) at low energies. The reasons for this are not yet
well understood (Weaver \& Gelbord 1999, in preparation). A
current pragmatic approach is to allow excess absorption in the SIS
detectors to account for this effect to zeroth order.  

Another source of low energy complexity is the partially ionized
absorber, first seen in the ROSAT PSPC spectrum (Madejski et al. 1995).  
A previous ASCA observation has shown that in IC4329a this complex
absorption is better modeled by two edges (corresponding to H and
He--like Oxygen at rest energies of 0.739 and 0.871 keV, respectively) rather
than a full ionized absorber code (Cappi et al. 1996; Reynolds 1997
see also the discussion in George et al. 1998). We
use this description here, but we caution that the determination of this
absorption in our data will be somewhat dependent on the way the low energy
calibration problems are treated.

We first use a
phenomenological model for the ASCA data, consisting of an underlying
power law and its reflected continuum from a neutral disk ({\tt
pexrav}: Magdziarz \& Zdziarski 1995) 
inclined at $30^\circ$, together with a separate Gaussian
iron line, with the two edge description for the warm absorber. 
This model gives a good fit to the data ($\chi^2_\nu=2533/2392$), 
for an intrinsic power law spectrum of $\Gamma=1.85\pm 0.03$, 
and reflector solid angle (for an inclination of
$30^\circ$) of $\Omega/2\pi=0.48_{-0.31}^{+0.34}$. 
The associated iron K$\alpha$
fluorescence line is at a (rest frame) energy of $6.37\pm 0.06$ keV, 
with equivalent width of $180\pm 50$ eV and intrinsic width of
$\sigma = 0.39\pm 0.10$ keV (hereafter all intrinsic line widths are 
the gaussian $\sigma$).

The iron line physical width is similar to that seen in a previous 
observation of this 
AGN (Mushotzky et al. 1995; Cappi et al. 1996; Reynolds 1997), but is
much smaller than that seen from the archetypal
relativistically smeared line in MCG--6--30--15 (Tanaka et al. 1995).
One way to show this is to fit the spectrum in the 2.5--10 keV band (where the
effects of the ionised absorption is much less) with a simple power law and its
expected reflected continuum (with solar abundances, and fixed solid angle
$\Omega/2\pi=1$ at $30^\circ$ inclination). The fit excludes the 5--7 keV iron
line range. Figure 2a shows the resulting shape of the line residuals from
IC4329a, while figure 2b shows those from MCG--6--30--15 for comparison.

\begin{figure*}
\plottwo{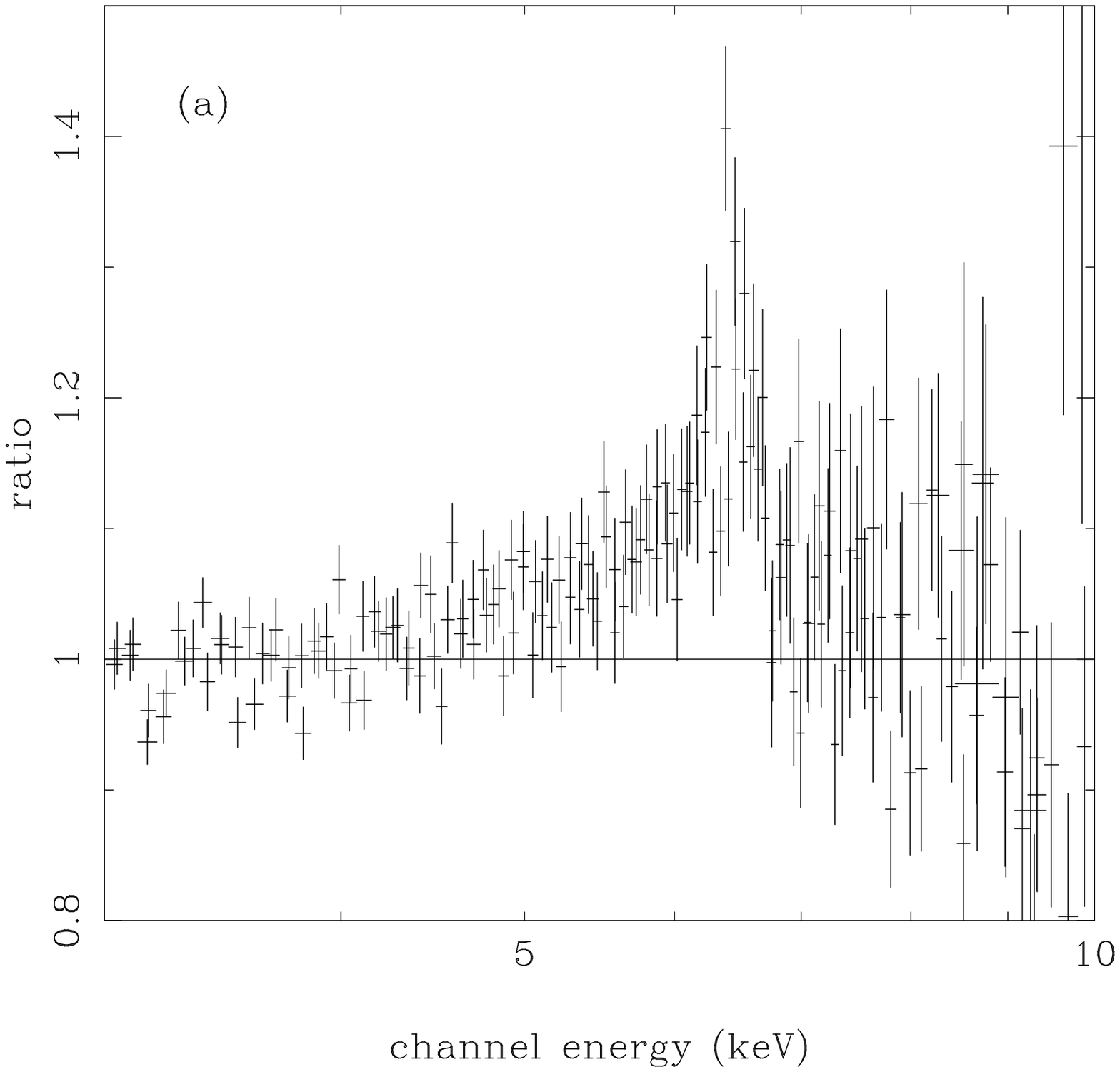}{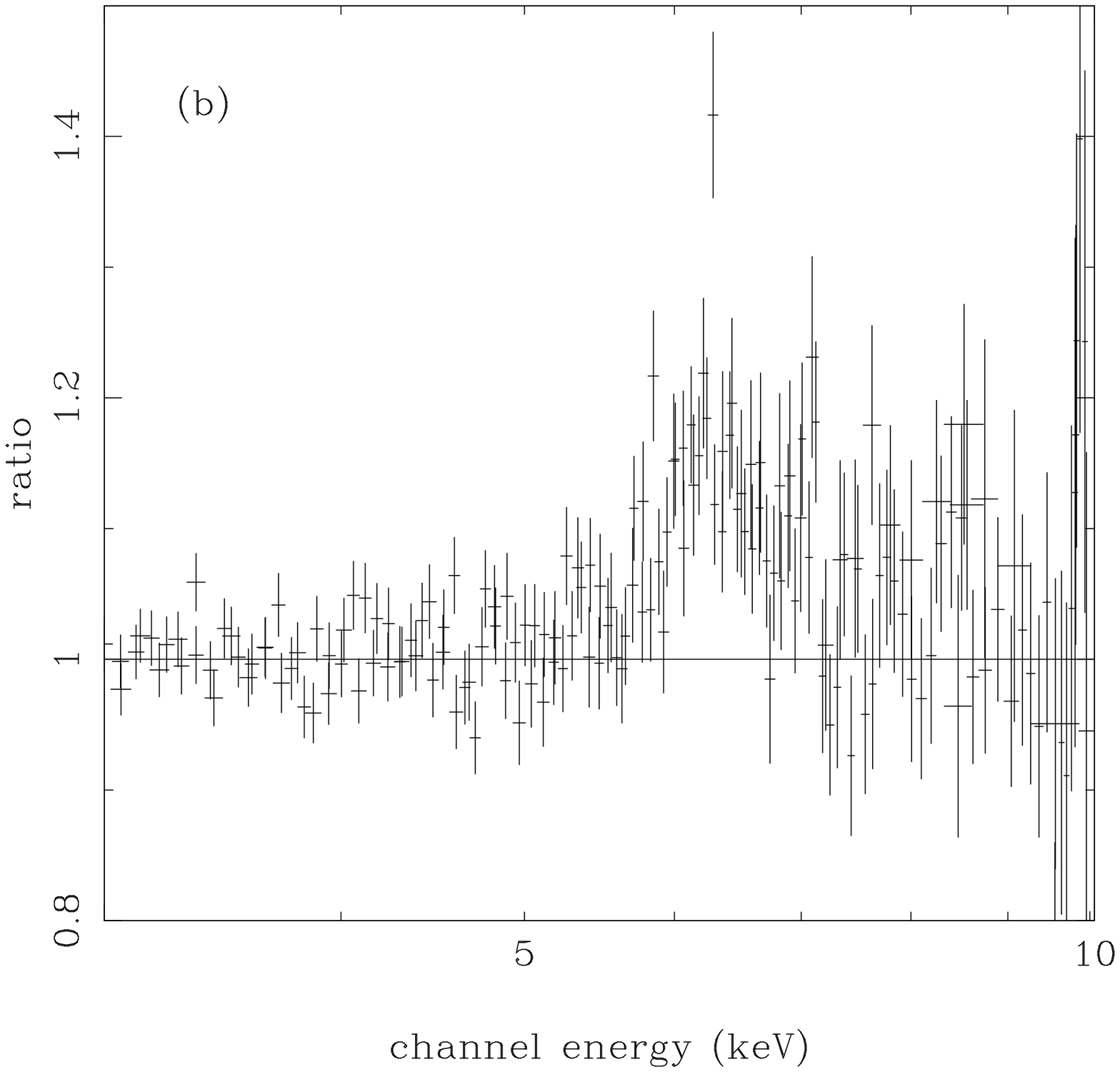}
\figcaption{Residuals of a continuum fit to a power law plus its Compton
reflected component over the 2.5--10 keV bandpass, excluding the iron line range
of 5--7 keV. (a) shows the residuals to the ASCA IC4329a data presented here,
while (b) shows residuals to the ASCA long look observation of MCG--6--30--15 of
Tanaka et al (1995). The line is clearly broader and skewed towards
lower energies in MCG--6--30--15 than in IC4329a.
}
\end{figure*}

Replacing the Gaussian line with a {\tt diskline} relativistic model (fixing the
disk inclination at $30^\circ$ and line emissivity at $\propto r^{-\beta}$ with
$\beta=3$) gives that the smearing implies an inner disk radius of
$48_{-20}^{+33} \Rg$ ($\chi^2_\nu=2540/2392$: a slightly worse fit than the
simple broad Gaussian line). This is significantly larger than the last stable
orbit at $6R_g$ ($\Rg=GM/c^2$, so the last stable orbit at 3 Schwarzchild radii
is at $6\Rg$).  The derived radius is dependent on the form of the emissivity,
but fits to two separate observations of MCG--6--30--15 require $\beta=3.4_{-0.8}^{+1.3}$ and
$\beta=4.4_{-1.1}^{+3.0}$, respectively (Nandra et al 1997), showing that
the line emission is strongly weighted towards the innermost radii. We might
also expect this on theoretical arguments. The
energy emitted per unit area of a disk goes as $r^{-3}$, so this should
give the time averaged emissivity from a magnetic corona, as well as being a
good approximation to the illumination expected from a central spherical source
(see appendix A in \.{Z}ycki\ et al 1999). Thus we fix $\beta=3$ in all our 
fits,
which gives $\Delta\chi^2 > 20$ for an inner radius equal to the last stable
orbit at {\it any} inclination.
 
The reflection description used above has several drawbacks. Firstly it allows
the line to vary in intensity (and energy) without reference to the reflected
continuum.  Secondly, the relativistic smearing is only applied to the iron
line, and not to the reflected continuum also.  We replace these components with
the reflection model described by \.{Z}ycki et al. (1999), {\tt rel-repr}, which
calculates the self--consistent line associated with the reflected continuum,
and then applies the relativistic smearing (including gravitational light
bending) to this total reprocessed spectrum. The ionisation state is a free
parameter, and the models are calculated for iron abundance between $1-2\times$
solar.  The reprocessor in AGN is generally assumed to be neutral but an
accretion disk temperature of $\sim 10$ eV can give a thermal population
dominated by ions with ionization potential $\sim 10-30 \times kT$ (Rybicki \&
Lightman 1979) i.e. between Fe V -- FeX, which in our model corresponds to $0.3
< \xi < 20$. Photo--ionisation by the X--ray source can increase these estimates
considerably.

We replace the {\tt pexrav} and {\tt diskline} components with our
combined model for the reprocessed component. We allow the ionisation,
iron abundance, and inner disk radius to be fit parameters, for a fixed
inclination of $30^\circ$, and for a fixed line emissivity of $\propto r^{-3}$.
This again gave  a disk inner radius of $45_{-18}^{+32} \Rg$
($\chi^2_\nu=2540/2392$).

\subsection{ASCA--XTE Simultaneous Data}

We extracted RXTE data which were taken exactly simultaneously with
the ASCA observation (datasets 4,5,6,11,12,13,16,16--01,18,19 and 20),
giving a total PCA exposure of 13 ks. The corresponding HEXTE data
over this short time interval have very low signal to noise so do not
add any appreciable constraints. Figure 3a shows the residuals
resulting from an absorbed power law fit to the PCA data, showing the classic
signature of Compton reflection and its associated iron fluorescence
line ($\chi^2_\nu=162.3/42$).  Figure 3b shows residuals including a
broad gaussian line in the fit. Clearly there are still systematic
residuals, with a decrement at the expected energy of the iron edge
and a rise to higher energies ($\chi^2_\nu=50.5/39$).  Including a
reflected continuum component (the {\tt pexrav} model in xspec,
assuming solar abundances and inclination of $30^\circ$) gives
$\chi^2_\nu=11.7/38$, showing that the reflected {\it continuum }
component is significantly detected independently of the iron line.

\begin{figure*}
\plottwo{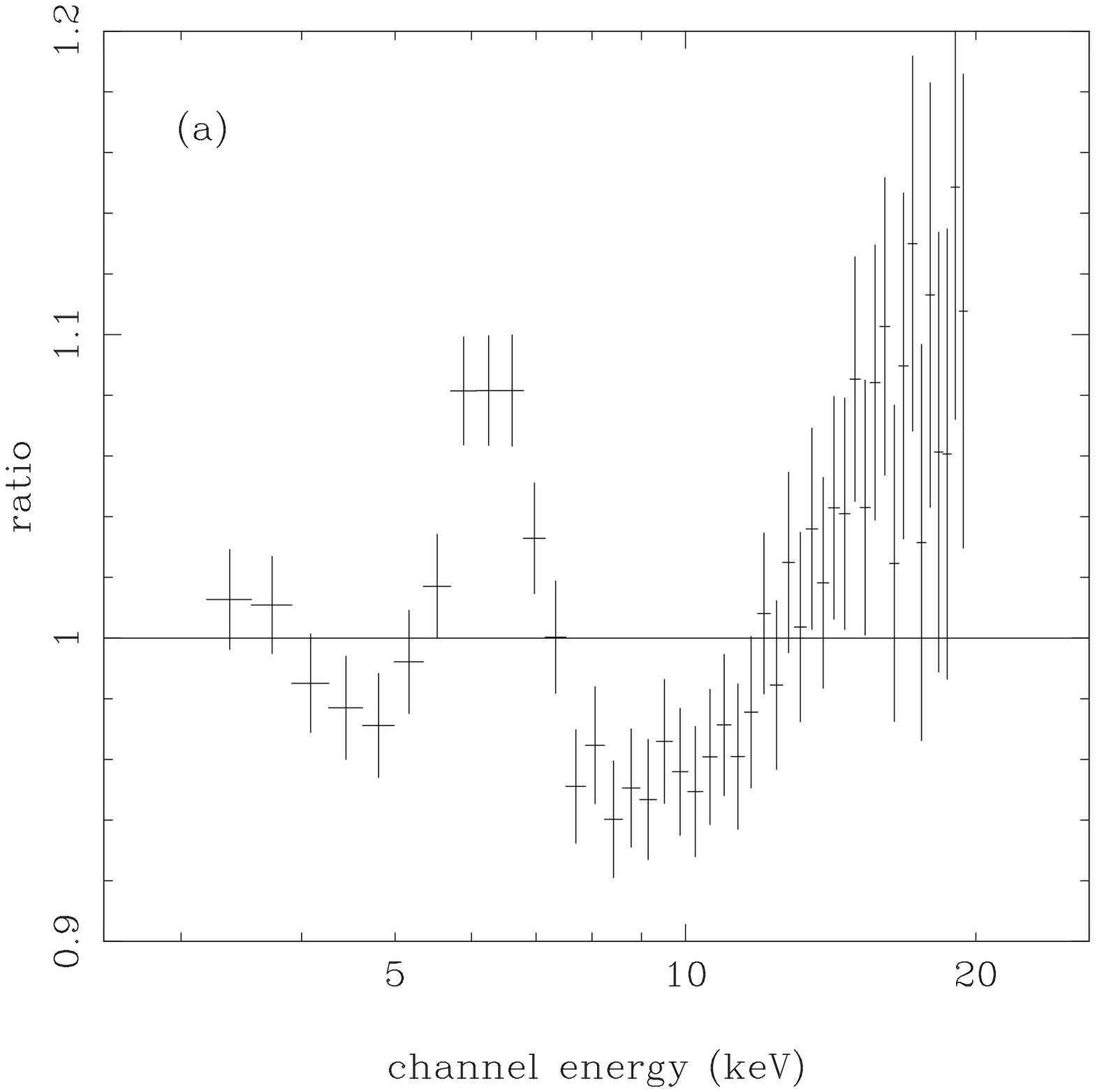}{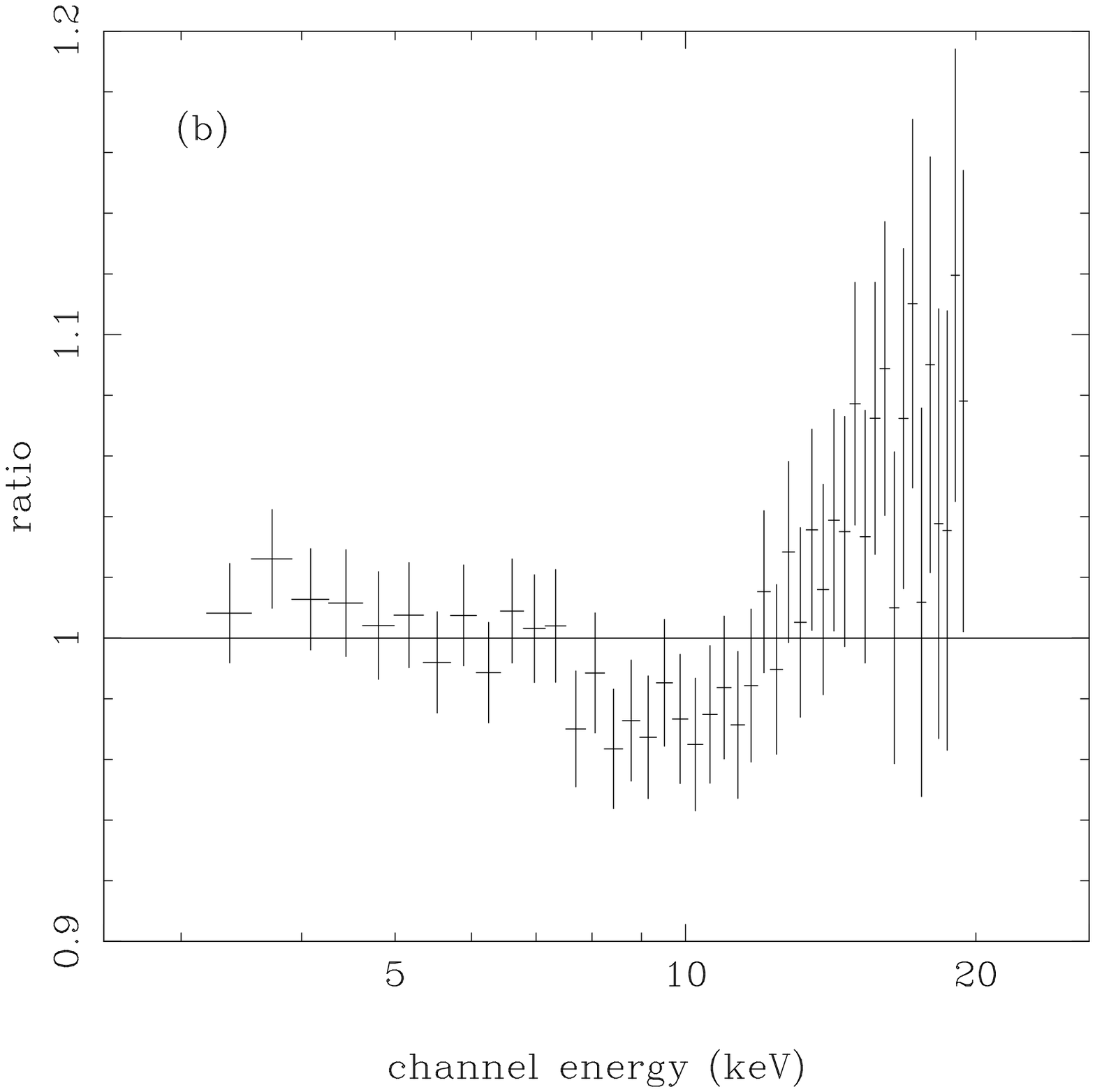}
\figcaption{XTE data: (a) shows residuals to a simple power law, showing the
classic reflection signature. (b) shows the remaining residuals including a
broad gaussian line. Plainly the reflected continuum is detected independently
of the iron line.
}
\end{figure*}

Since the data are simultaneous, we can 
fix the absorption (warm and cold) 
to that seen in the GIS for the same model fit to the ASCA data. This gives 
an excellent fit to the data with $\chi^2_\nu=12.6/39$
(rather too good in fact, showing that the
statistics are dominated by the 1 per cent systematic error)
for an intrinsic power law spectrum of
$\Gamma=1.94\pm 0.04$,
and reflector solid angle (for an inclination of
$30^\circ$) of 
$\Omega/2\pi=0.49^{+0.17}_{-0.14}$. The associated iron K$\alpha$
fluorescence line is at a (rest frame) energy of $6.33\pm 0.13$ keV,
with equivalent width of $210 \pm 45$ eV and intrinsic width (gaussian
$\sigma$) of
$0.49^{+0.19}_{-0.18}$ keV. The PCA 2--10 keV flux is 25 per cent higher than
that from ASCA due to absolute flux calibration problems in both instruments:
ASCA gives a Crab 2--10 keV flux of $1.8\times 10^{-8}$ ergs s$^{-1}$
(Makishima et al., 1996), while RXTE gives $2.4\times 10^{-8}$ 
ergs s$^{-1}$ (see PCA Crab spectrum at
http://lheawww.gsfc.nasa.gov/users/keith/pcarmf.html) 
and the original Crab calibration is $\sim 2.15\times 10^{-8}$ ergs s$^{-1}$ (Toor
\& Seward 1974). A more serious discrepancy is in the spectral index, which is
$\Delta\alpha\sim 0.1$ steeper than that seen in ASCA. 
It is known that the current RXTE PCA calibration gives
results for the Crab which are roughly $\Delta\alpha\sim 0.1$ steeper than the
index assumed for the calibration of other instruments (K. Jahoda, private
communication).  However, reassuringly, the relative amount of reflection
equivalent width, energy and intrinsic width of the line 
are consistent with those from the ASCA data, though of course the absolute
normalisations are different due to the calibration discrepancies.
Tieing the absorption across the two datasets, together with the 
relative reflection parameters (the reflector solid angle, its 
inner radius and ionisation state, and the iron line
energy, width and equivalent width)
gives $\chi^2_\nu=2548/2435$, not significantly different
from the $\chi^2_\nu=2546/2432$ obtained from the separate fits. Thus in what follows we
tie the relative reflection (and absorber) parameters across the two datasets, and 
let only the power law spectral index and normalization be free.

We replace the {\tt pexrav} and broad Gaussian line with our {\tt rel-repr}
model. Table 1 shows the results of a joint fit to the ASCA and RXTE PCA data
for inclination angles of $30$, $60$ and $72^\circ$ and for iron abundances of
$1$ and $2\times$ solar. 
The derived inner disk radius is highly correlated with the assumed inclination.
High inclination angles give stronger Doppler effects and so a broader line.
The inner disk radius then has to be larger to match the observed line width.
However, {\it none} of the fits allow the disk to extend down to the innermost
stable orbit, irrespective of inclination. This also means that 
the data are not very sensitive to the
inclination ($\Delta\chi^2\sim 4$, i.e. marginally significant preference for
higher inclinations). It is only the broadest components from the very innermost disk
which significantly change the skewness of the line profile (as opposed to its
width) as function of inclination.

The data {\it are} sensitive to the iron abundance. They significantly 
prefer supersolar abundances ($\Delta\chi^2\ge 7$), as would be expected from measurements of 
radial abundance gradients in spiral galaxies (e.g. Henry \& Worthey 1999). 
Another way to see this is
the phenomenological fits give a line equivalent with of $180$ eV, as expected
for a solar abundance slab subtending a solid angle of $2\pi$ (e.g. George \&
Fabian 1991), yet the solid angle of the reflected continuum in these fits is 
approximately half of this. Figure 4 shows the best fit joint RXTE and ASCA 
data fit to a disk model with $2\times$ solar abundance, inclined at $30^\circ$.

\begin{figure*}
\plotone{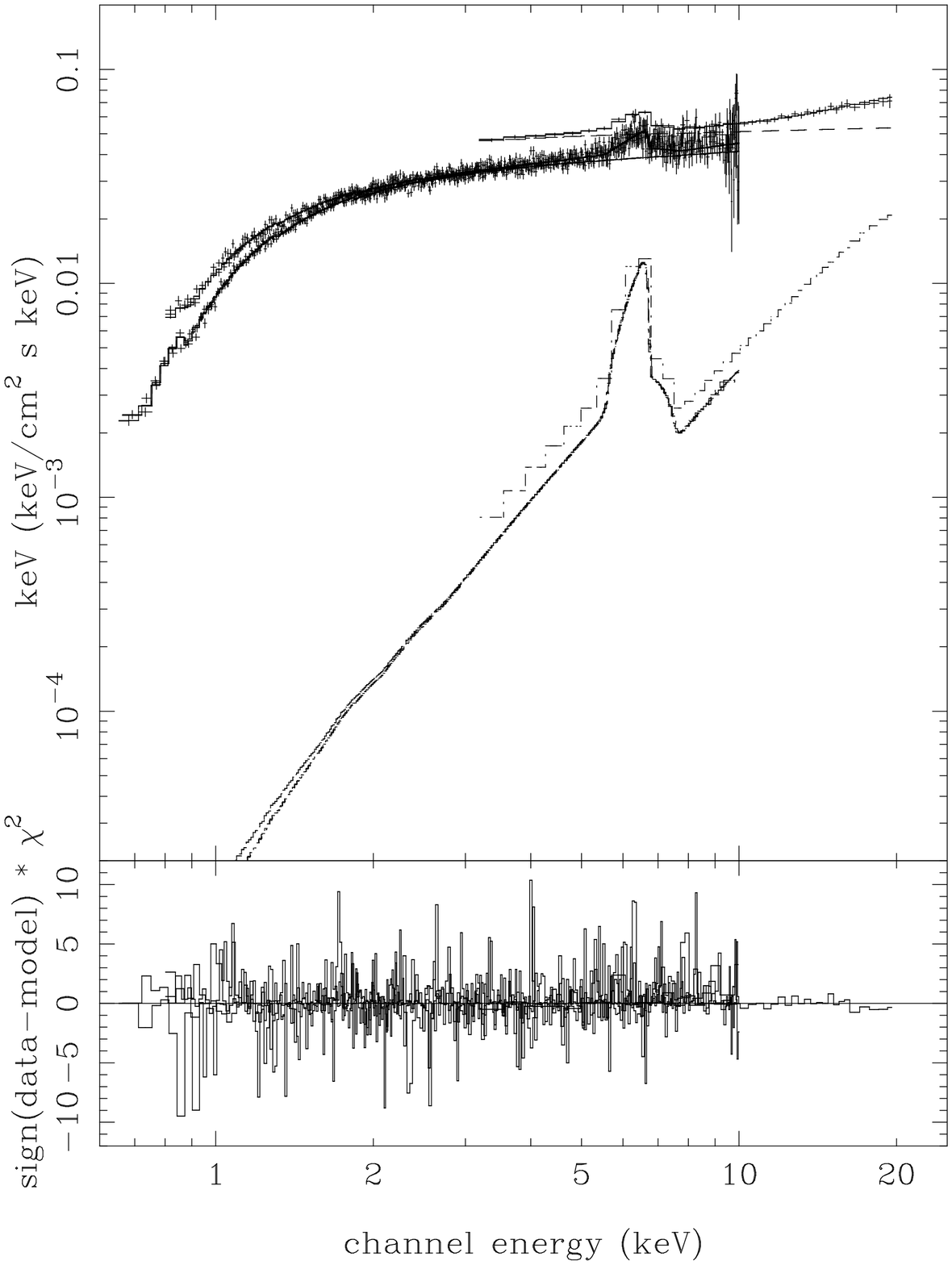}
\figcaption{ASCA and simultaneous RXTE data fit to a model in which the
relativistic reprocessor is assumed to have twice solar iron abundance, and is
inclined at $30^\circ$. The ASCA and RXTE intrinsic power law index is not tied
between the two instruments due to calibration uncertainties.}
\end{figure*}

\begin{deluxetable}{lccccccc}
 \footnotesize
 \tablewidth{0pt}
 \tablecaption{ASCA and simultaneous XTE PCA data fit to the {\tt rel-repr}
 model }
 \tablehead{
\colhead{${\rm A_{Fe}}$\tablenotemark{a}} &
                      \colhead{Inclination} &
                     \colhead{$\Gamma$\tablenotemark{b}} &
                     \colhead{Norm at 1 keV} & 
                     \colhead{$\Omega/2\pi$\tablenotemark{c}} &
                     \colhead{$\xi$\tablenotemark{d} }  &
                     \colhead{$\Rin$\tablenotemark{e}\ ($\Rg$)} &
\colhead{$\chi^2_\nu$} \nl
& & \colhead{ASCA, XTE} & \colhead{ASCA, XTE}\nl
}
\startdata

1 & 30 & $1.87\pm 0.02$, $1.99_{-0.04}^{+0.03}$ & $0.031, 0.049$ & $0.62^{+0.11}_{-0.13}$
  & $21_{-13}^{+45}$ & $35_{-15}^{+20}$ & 2567/2436 \nl
1 & 60 & $1.88\pm 0.02$, $2.00_{-0.02}^{+0.04}$ & $0.031, 0.05$ & $0.85^{+0.16}_{-0.10}$
  & $12^{+10}_{-8}$ & $90^{+60}_{-30}$ & 2575/2436 \nl
1 & 72 & $1.89\pm 0.02$, $2.02\pm 0.04$ & $0.031, 0.051$ & $1.25^{+0.23}_{-0.20}$ 
  & $7^{+10}_{-5}$ & $110^{+80}_{-38}$ & 2576/2436 \nl
2 & 30 & $1.85\pm 0.02$, $1.94_{-0.02}^{+0.03}$ & $0.030, 0.046$ & $0.55\pm 0.10$ 
  & $4_{-3.95}^{+11}$ & $35^{+16}_{-12}$ & 2562/2436 \nl
2 & 60 & $1.86_{-0.02}^{+0.01}$, $1.97\pm 0.02$ & $0.030, 0.048$ & $0.88\pm 0.11$ 
  & $0.01_{-0.01}^{+0.8}$ & $75_{-23}^{+45}$ & 2560/2436 \nl
2 & 72 & $1.86\pm 0.02$, $1.95^{+0.03}_{-0.02}$ & $0.030, 0.048$ & $1.20^{+0.13}_{-0.10}$
  & $0.005_{-0.005}^{+0.3}$ & $90_{-25}^{+40}$ & 2558/2436 \nl

\enddata
\tablenotetext{a}{Iron abundance of reflector relative to solar}
\tablenotetext{b}{Photon spectral index}
\tablenotetext{c}{Solid angle subtended by the reprocessor with respect to the
X--ray source}
\tablenotetext{d}{Ionization parameter of the reprocessor}
\tablenotetext{e}{Inner radius of the accretion disk}

\end{deluxetable}

Since IC4329a is a Seyfert 1 there could be a contribution
to the line/reflected continuum from 
a molecular torus as well as from the accretion disk (Ghisellini, Haardt \& 
Matt 1994; Krolik, Madau \& \.{Z}ycki 1994). 
Table 2 shows the results obtained including a
neutral, unsmeared reflector, with assumed mean inclination of $60^\circ$.
Only the parameters of the two reflectors are included, since the continuum
is similar to that derived before. This gives a {\it significantly} better fit to the data,
generally with $\Delta\chi^2\ge 9$ for the addition of 1 extra free parameter
(the amount of non--relativistic reflection). 

The double reprocessing model now gives a fit which is as good as or even
better than those from the phenomenological (i.e. unphysical) broad Gaussian
line/{\tt pexrav} model. The observed line is fairly broad but also fairly
symmetric, contrary to the predictions of a line from an accretion disk which
{\it must} also be skewed if it is broad.  Adding the second neutral,
non--smeared reprocessor gives a narrow core to the line, while the line
from the relativistic reprocessor then fills in a broad (and skewed) line
wing. The presence of the narrow component means that the relativistic effects
have to be more marked than in the single reflector fits in order to make the
total line as broad as before.  Thus the derived
inner disk radius is always rather smaller than before {\it but still never
consistent with the 3 Schwarzschild radii}. The smaller inner radius means that
the lower inclination reflected spectra are significantly gravitationally
redshifted, so requiring the reflector to be somewhat ionized to compensate for
this.

The above discussion assumed that the torus was Compton thick,
i.e. with $N_H \gg 10^{24}$ cm$^{-2}$. However, it can still produce
substantial line emission, without the accompanying strong reflected
component if the torus column is $\sim 10^{23-24}$
cm$^{-2}$. Replacing the unsmeared, neutral reflected component by a
narrow 6.4 keV line gives a similar series of fits as those shown in Table
2. In particular, the fits {\it never} allow the amount of smearing to
be as large as expected from the innermost stable orbit of an
accretion disk, and they show the same preference for twice solar iron
abundance and inclination angles $> 30^\circ$.

\begin{deluxetable}{lccccccc}
 \footnotesize
 \tablewidth{0pt}
 \tablecaption{ASCA and simultaneous XTE PCA data fit to the {\tt rel-repr}
 model, with relativistic disk and neutral, unsmeared reflection }
 \tablehead{
\colhead{${\rm A_{Fe}}$\tablenotemark{a}} &
                      \colhead{Inclination} &
                     \colhead{$\Omega/2\pi$\tablenotemark{b}} &
                     \colhead{$\Omega/2\pi$\tablenotemark{c}} &
                     \colhead{$\xi$\tablenotemark{d} }  &
                     \colhead{$\Rin$\tablenotemark{e}\ ($\Rg$)} &
\colhead{$\chi^2_\nu$} \nl
}
\startdata

1 & 30 & $0.40\pm 0.14$ & $0.25_{-0.07}^{+0.09}$ & $120_{-50}^{+110}$ 
  & $18_{-6}^{+10}$ & 2550/2435\nl
1 & 60 & $0.52_{-0.14}^{+0.13}$ & $0.33_{-0.09}^{+0.14}$ & 
$100_{-55}^{+90}$ 
  & $42_{-18}^{+25}$ & 2551/2435\nl
1 & 72 & $0.53_{-0.14}^{+0.13}$ & $0.46_{-0.14}^{+0.24}$ & $90_{-55}^{+70}$ 
  & $55^{+30}_{-24}$ & 2551/2435\nl
2 & 30 & $0.29\pm 0.11$ & $0.25_{-0.09}^{+0.15}$ 
  & $60^{+100}_{-50}$ & $16_{-5}^{+10}$ & 2552/2435 \nl
2 & 60 & $0.33_{-0.11}^{+0.10}$ & $0.51^{+0.23}_{-0.18}$ & $5_{-5}^{+40}$ 
  & $32_{-10}^{+21}$ & 2545/2435 \nl
2 & 72 & $0.31\pm 0.12$ & $0.85_{-0.35}^{+0.21}$ & $0.06_{-0.06}^{+25}$ 
  & $40_{-12}^{+28}$ & 2544/2435 \nl

\enddata
\tablenotetext{a}{Iron abundance of reflector relative to solar}
\tablenotetext{b}{Solid angle subtended by the neutral, unsmeared reprocessor
(assumed mean inclination of $60^\circ$) with respect to the
X--ray source}
\tablenotetext{c}{Solid angle subtended by the relativistic reprocessor with respect to the
X--ray source}
\tablenotetext{d}{Ionization parameter of the relativistic reprocessor}
\tablenotetext{e}{Inner radius of the relativistic accretion disk}

\end{deluxetable}

The series of fits above show that the best physical description of the data is
with two reprocessed components, one which is relativistically smeared and
possibly ionized from the accretion disk, and one which is neutral and 
unsmeared (and possibly consisting of just line rather than line and 
reflected continuum)
from the molecular torus. A similar composite line is seen in the Seyfert
MCG--5--23-16 (Weaver et al.\ 1997; see also Weaver \& Reynolds 1998). 
The data prefer 
models with twice solar abundances, and inclinations of $> 30^\circ$,
and these solutions have the advantage that the
derived ionisation of reflector is generally rather low, consistent
with that expected (see section 2). However, there is a further constraint
on the inclination, since the extra reprocessor cannot be along our line
of sight to the nucleus. IC4329a is classified optically as a Seyfert 1
and the X--ray spectrum is not heavily absorbed. 
For a disk inclined at $72^\circ$ then for our line of sight not to
intercept the molecular torus severely restricts its scale height, and so the
possible solid angle it can subtend. Thus while the range of double reflector
fits given in Table 2 are statistically indistinguishable, these consistency
arguments lead us to favor viewing angles to the accretion disk of $\sim
45^\circ$ and supersolar abundances.

All these models assumed that the ionisation state of the disk was
constant with radius. A more physical picture might be one where the
ionisation varies as a function of radius (e.g. Matt et al. 1993). In
such models, the inner disk might be so ionized that it produces no
spectral features. The observed reflected spectrum would then arise
from further out in the disk, and so not contain the highly smeared
components.  This might provide an alternative explanation to a
truncated disk as to why the relativistic smearing observed is not
compatible with a disk extending down to the last stable orbit. We
test this by dividing the disk into 10 radial zones, with ionisation 
$\xi(r)\propto r^{-4}$ as described in Done \& \.{Z}ycki (1999). 
With the inner radius fixed at $6\Rg$, and allowing for a narrow component from 
the molecular torus we are never able to obtain fits within 
$\Delta\chi^2\sim 20$ of those in Table 2.

\subsection{RXTE PCA Variability}

There is clear intensity variability during the RXTE campaign, and also
spectral variability in the sense that the spectrum becomes softer as
the source brightens. This could be due to either
intrinsic change in the power law spectral index, or a change in the
relative contribution of the reflected spectrum (or both).  

We can attempt to disentangle the power law from the reflected spectrum
by fitting these components to the individual spectra from each orbit
(including a 1 per cent systematic uncertainty), fixing the absorption
at $3\times 10^{21}$ cm$^{-2}$.  The reflected spectrum is assumed to
have twice solar iron abundance, be inclined at $60^\circ$ and have
fixed negligible ionisation ($\xi=0.01$, see table 1).  Thus the free
parameters are the power law index and normalization, the solid angle
subtended by the relativistic reflector and its inner radius.  Figure
5a shows these derived parameters as a function of time.  Clearly the
inner radius cannot be constrained, but the index and the solid angle
of the reflector show some possible trend. Fitting these with
a constant gives $\chi^2_\nu=36.0/58$ and $13.4/58$ respectively
i.e. they are statistically consistent with a constant value.  Figure
5b shows these plotted against the 2--10 keV flux, and a linear
regression (taking errors in both $x$ and $y$ into account: Press et
al.\ 1992) shows that the power law index is {\it significantly}
correlated with flux, since it gives $\chi^2_\nu=25.7/57$. 
This corresponds to an F value of $10.3/(25.7/57)= 22.8$, 
significant at $\ge 99.9$ per cent. 
Even using just the difference in $\chi^2$ between the two fits 
gives $F=10.3$, significant 
at $\ge 99.5$ per cent, so the correlation is clearly
present. IC4329a then becomes only the second Seyfert 1 where there is
clear {\it intrinsic} spectral variability (the other is NGC 5548:
Magdziarz et al. 1998), where underlying continuum changes can be
unambiguously disentangled from changes in the reflected spectrum. 

We use the same procedure to look for variability in the amount of reflection
as a function of flux. The linear regression
gives $\chi^2_\nu=10.6/57$ 
where the fractional amount of reflection relative to the
power law {\it decreases} as the flux increases. 
This is significant at $> 99.9$ per cent confidence on an F test,
but is only 90 per cent significant using just the change in 
$\chi^2$.

\begin{figure*}
\plottwo{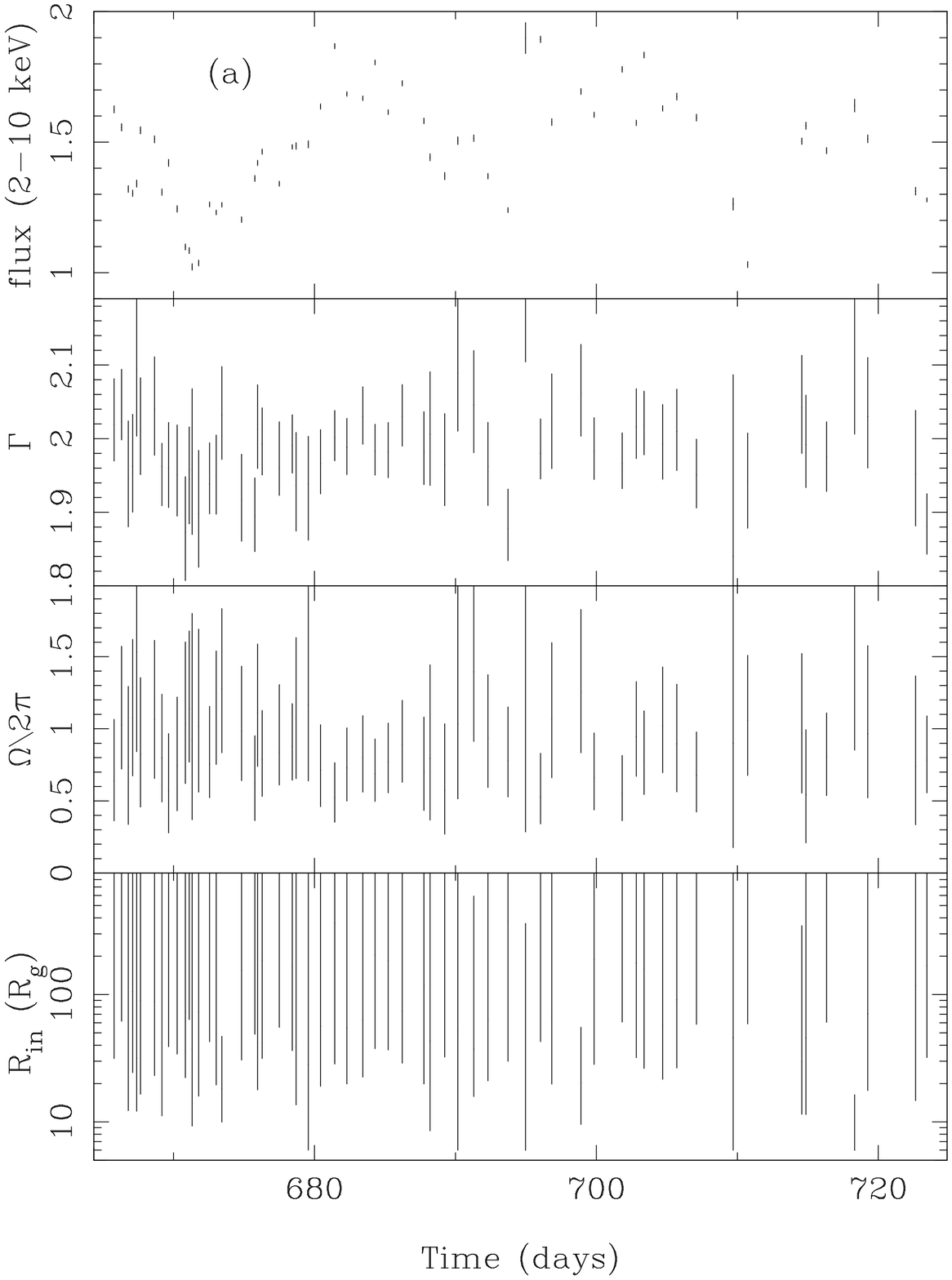}{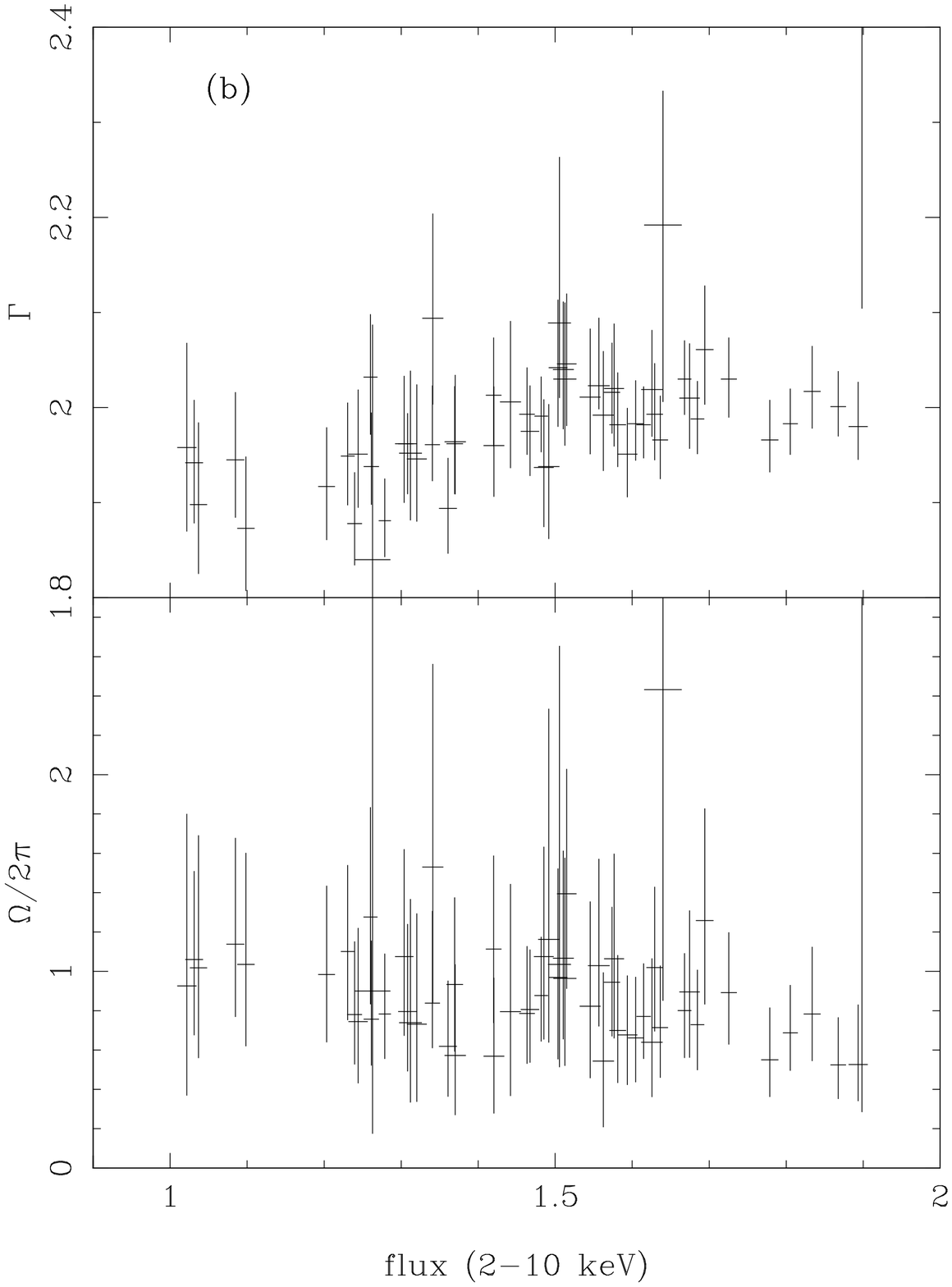} 
\figcaption{Results for
fitting a simple power law and its relativistic reflection with
absorption fixed at $3\times 10^{21}$ cm$^{-2}$ to the 3--20 keV
individual snapshot PCA spectra.  (a) shows the inferred 2--10 keV
flux, spectral index and derived inner disk radius as a function of
time, while (b) shows the spectral index and amount of reflection as a
function of 2--10 keV flux. The power law becomes intrinsically
steeper as the source brightens, while the reflected fraction marginally
decreases.}
\end{figure*}

To illustrate these points we co--add spectra when the source was at
its lowest and highest intensity level (see Figure 1), and fit these
spectra with a power law and single relativistically smeared reflector
(with reflector parameters fixed as above, and with the disk inner
radius fixed at $60\Rg$). The spectral index changes from $1.92\pm 0.04$
to $2.00\pm 0.02$, for the low
and high state, respectively. Figure 6 shows the unfolded spectra, with
the model extrapolated out to 100 keV. The thick and thin lines show
the model components for the high and low state data, respectively.
The underlying continuum is brighter at all energies $\le 100$ keV 
in the high state despite it being steeper, with an integrated 0.01--300 keV
flux of $6.2$ to $10.7 \times 10^{-10}$ ergs cm$^{-2}$ s$^{-1}$ for
the low and high state, respectively. This behaviour is fairly
easy to reproduce in comptonisation models in which the seed photons
vary (see discussion), but the (marginally significant) 
lack of change in the absolute amount
of reflection (so that the relative reflected fraction in the low
state is larger than in the high state) is harder to explain.
Clearly a model in which the reflected flux is produced at large
distances from the source would be viable, but the
reflected spectrum is broadened,
so does contain at least some contribution from the relativistically
smeared inner disk. For a $10^8\msun$ black hole, then the inner 100
Schwarzschild radii are on scales of $3\times 10^{15}$ cm, i.e. less
than 1 lightday. The spectra are taken at intervals of a day or more,
so the relativistically smeared reprocessed component should not be
appreciably lagged behind the source variability.

\begin{figure*}
\plotone{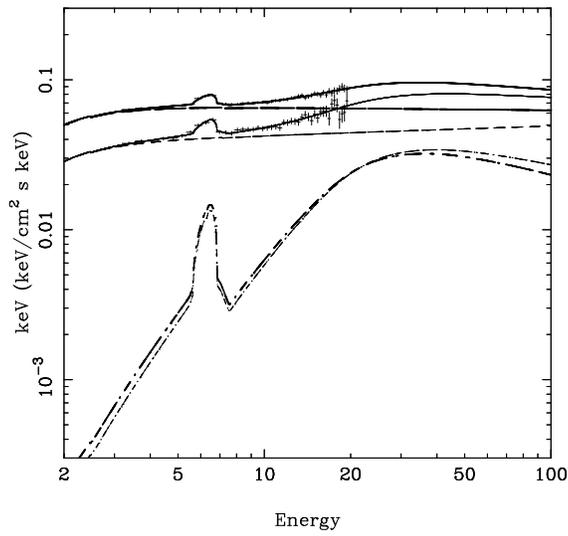}
\figcaption{The high and low state PCA spectra fit to a simple power
law and its relativistic reflection with absorption fixed at $3\times
10^{21}$ cm$^{-2}$. The best fit model components are shown in thick
and thin lines for the high and low state spectra, respectively, where
the spectral index and relative amount of reflection are $\alpha=
2.00\pm 0.02$, $1.92\pm 0.04$, and $\Omega/2\pi=
0.74\pm 0.13$, $1.07^{+0.25}_{-0.22}$.  
The model is extrapolated out to 100 keV, showing that the high
state power law has higher intensity at all energies of importance for
the formation of the reflected continuum, despite being
steeper (the pivot energy is at $\sim 1$ MeV).
}
\end{figure*}

One possibility is that this anti--correlation of relative reflected
fraction with flux is intrinsic to the
source, that the geometry changes in such a way as to produce less
reflection as the source brightens and steepens. However, it is very
hard to see how this can be the case.  A brighter, steeper source
implies that there are more seed photons for the Compton scattering
(see Figure 6 and the Discussion),
i.e. that the geometry is such that more disk photons are intercepted
by the source. Thus we can easily explain more reflected flux as the
source steepens, but not less. A {\it correlation} of reflected solid
angle with spectral index is indeed seen in both Galactic Black Hole
Candidates (e.g. \.{Z}ycki et al. 1999) and AGN (Zdziarski et al. 1999). It
seems much more likely that the (marginal) {\it anti--correlation} is an
artifact of there being a second reflector at much larger distances
from the source.  The light travel time delay then means that the
distant reprocessor does not have time to respond to rapid flux
variations, so that as the source dims the relative contribution of
the reflected spectrum from the distant reprocessor increases.

We include a reprocessed component from a torus, fixing its parameters
to the best fit XTE values from the joint ASCA--RXTE fit (see table 2,
again assuming twice solar abundance and inclination of $60^\circ$,
but this time also fixing the inner edge of the disk to $60 \Rg$). The
significance of the correlation of the spectral index with flux is
unchanged ($\chi^2_\nu=44.7/58$ for a constant while adding the linear
term gives $\chi^2_\nu=29.4/57$), while the flux/accretion disk
reflection variability is now consistent with a constant solid angle
($\chi^2_\nu=12.4/58$): adding a linear term gives an insignificant
change ($\chi^2_\nu=11.4/58$).  The results are similar for a fixed
Gaussian line rather than a full reprocessed spectrum.

We illustrate this again by fits to the high and low state spectra.
Figure 7a shows the confidence contours for the power law spectral
index, while Figure 7b shows the derived solid angle of the
relativistic reflector. The diagonal lines denote solutions where the
power law index and reflected fraction remain constant between the two
datasets. The power law index is clearly variable between the high and
low flux level datasets irrespective of how the reprocessor is
modelled. With a single reprocessor then the relative amount of
reflection is only consistent with remaining constant at the $<90$ per
cent confidence level. The data prefer a larger contribution of
reflected flux relative to the power law in the low state spectrum
i.e. that the absolute normalization of the reflected flux is
constant. The dashed and dotted lines show the same contours for a
model including an unsmeared, neutral reprocessed spectrum and
Gaussian line from the molecular torus, respectively. The reflected
fraction is then consistent with a constant value.

\begin{figure*}
\plottwo{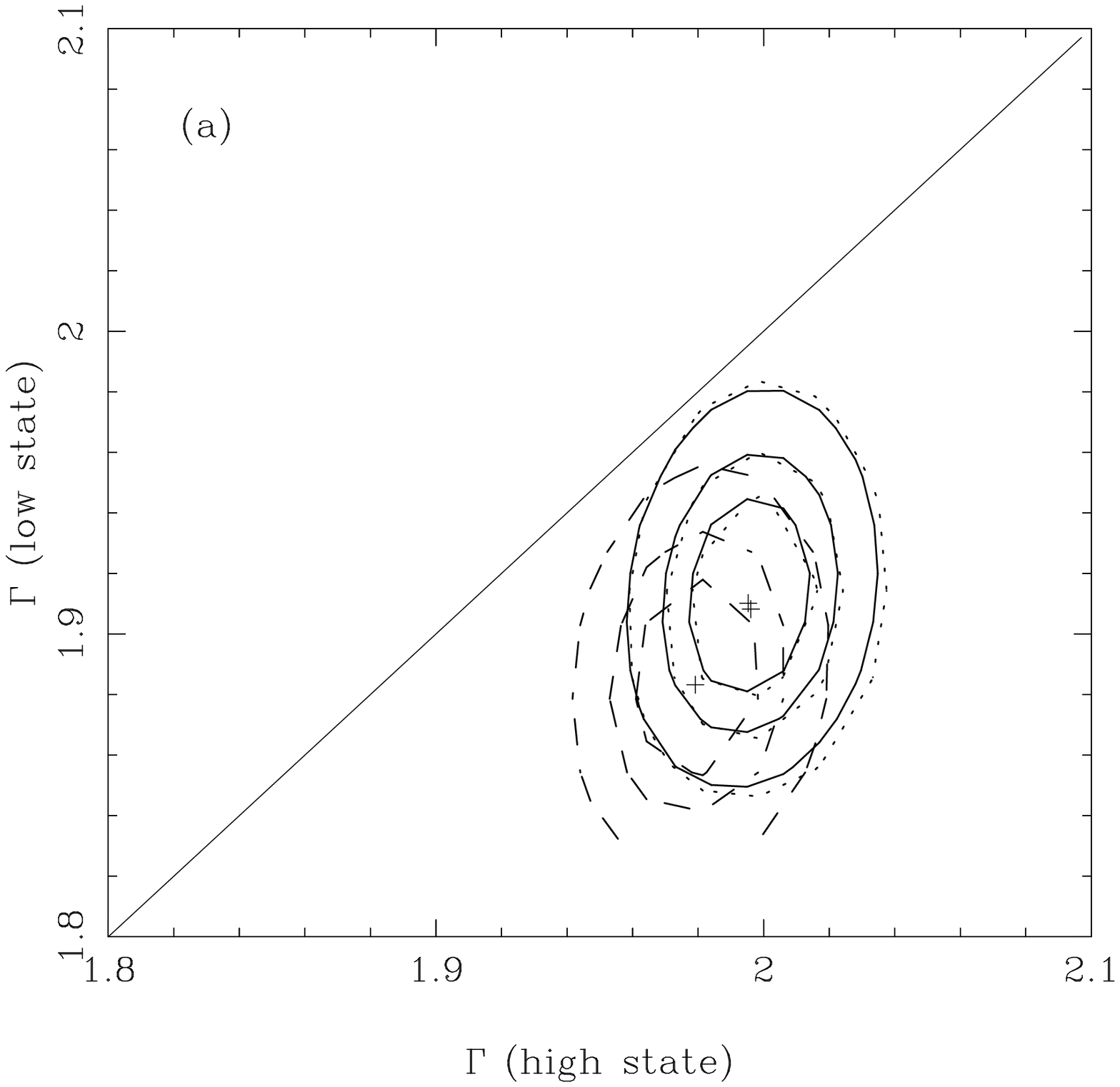}{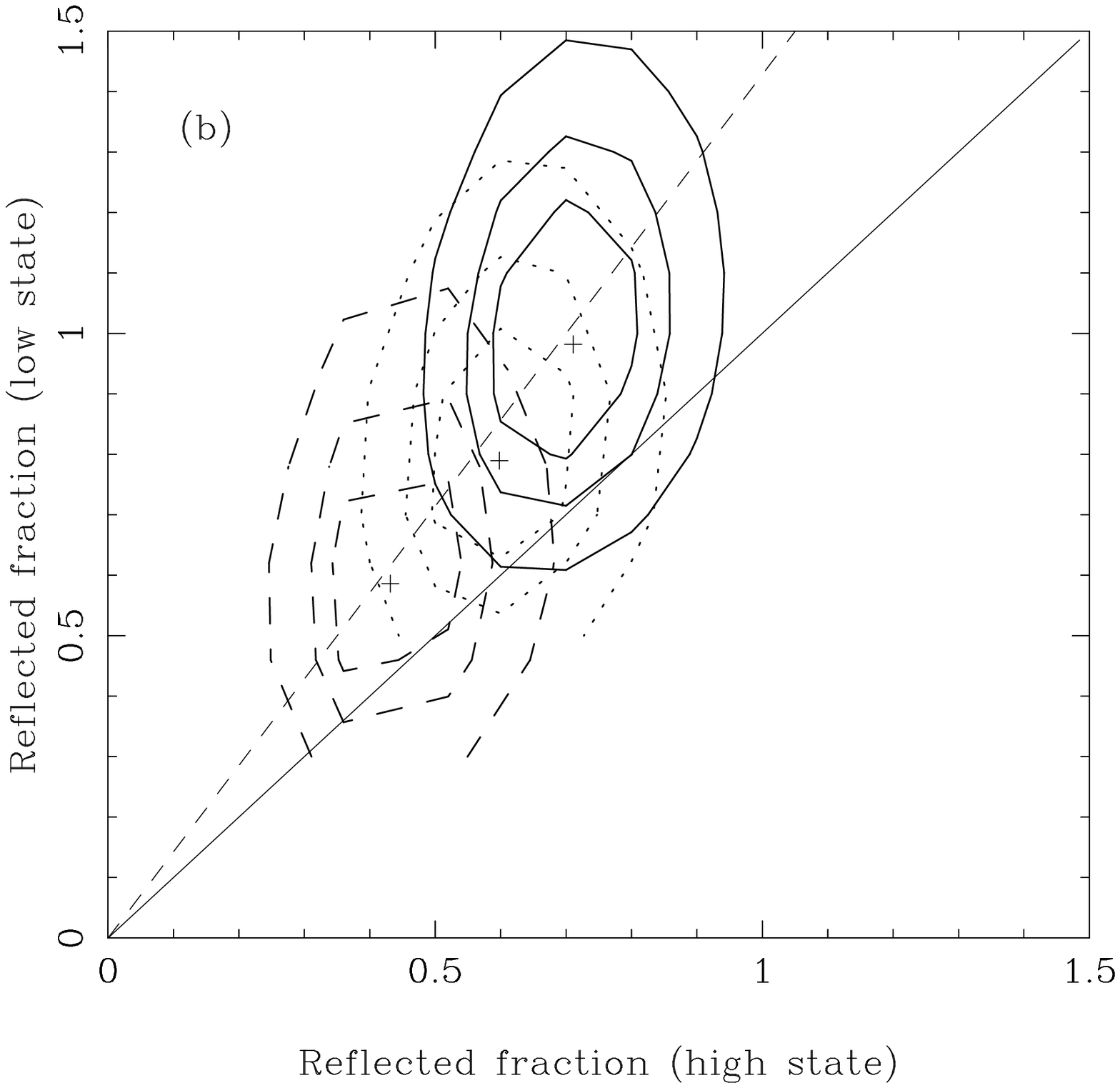}
\figcaption{Confidence contours for (a) the power law spectral index
and (b) the reflected fraction in the high and low state spectra.
Solid contours denote a spectral model where the reprocessing is from
an accretion disk, while the dashed contours include a fixed
contribution from reflection from a molecular torus and the dotted
contours include a fixed Gaussian line. In (a) the solid straight line
denotes constant spectral index.  This is plainly outside all the
model contours, strongly requiring that the intrinsic power law index
changes. In (b) the solid straight line indicates a constant reflected
fraction, while the dashed line shows a constant absolute
normalisation of the reflected component.  Modeling the accretion disk
alone gives confidence contours which are only marginally consistent
at the 90 per cent confidence level with a constant reflected
fraction.  Instead the data prefer a constant absolute normalisation
of the reflected component.  Including a constant component (line or
full reprocessed spectrum) shifts the derived contours for the
accretion disk, allowing a constant reflected fraction at higher
confidence level. }
\end{figure*}

\subsection{PCA, HEXTE and OSSE Total Spectrum}

All the RXTE PCA data were co--added to form a single spectrum (with 1
per cent systematic uncertainty added) and fit together with the HEXTE
and OSSE data to give a broad band spectrum. Current consensus is that
the continuum is formed by Compton scattering of soft seed photons by
hot electrons. Such a Comptonised continuum can be approximated by a
power law with exponential cutoff, but this becomes inaccurate if the
spectrum extends close to the energies of either the seed photons or
hot electrons. The seed photons are presumably from the accretion
disk, with expected temperatures of $\sim 10$ eV, so the spectral
curvature here is not an issue. However, the inclusion of the OSSE
data means that the shape of the spectral cutoff from the electron
temperature becomes important. Thus we use an analytical approximation
to a Comptonised spectrum based on solutions of the Kompaneets
equation (Lightman \& Zdziarski 1987), where the shape of the high
energy cutoff is sharper than an exponential rollover.

The relativistic reflection and line from this incident continuum are
calculated, and included in the model fit. The results are detailed in
Table 3 for inclinations of the relativistic reprocessor of $30, 60$
and $72^\circ$ and iron abundances of $1$ and $2\times$ solar.  We see
the same trend as in Table 1 in terms of the spectra preferring higher
iron abundance, and these solutions constrain the electron temperature
to be $kT_e\sim 40-100$ keV (corresponding to an exponential e-folding
energy of $\sim 120-300$ keV).  Figure 8 shows the best fit solution
for twice solar abundance, inclined at $60^\circ$.  However, the poor
signal--to--noise of the data at the highest energies means that this
temperature is dependent on details of the reflection spectrum. For
solar abundances the temperature is generally unconstrained.

\begin{figure*}
\plotone{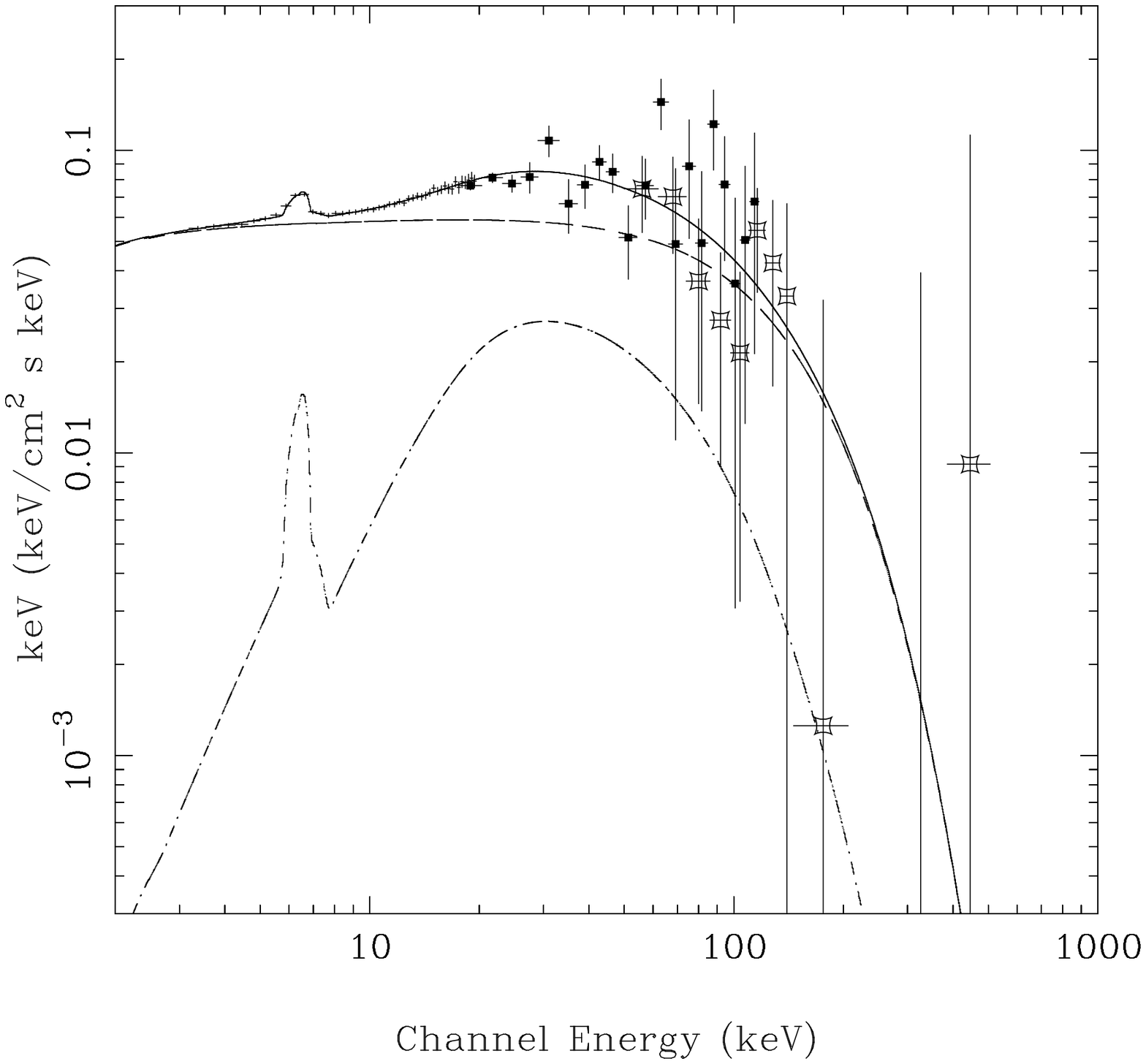}
\figcaption{The total PCA (crosses), HEXTE (filled squares) and OSSE 
(open squares) spectra, fit to a comptonised
continuum model with $kT_e\sim 50$ keV, together with its relativistic
reflection component for an assumed inclination of $60^\circ$ and twice solar
iron abundance}
\end{figure*}

\begin{deluxetable}{lccccccccc}
 \footnotesize
 \tablewidth{0pt}
 \tablecaption{Total XTE PCA, RXTE HEXTE and CGRO OSSE data fit to an approximate
 comptonised continuum and its relativistically smeared reflection}
 \tablehead{
\colhead{${\rm A_{Fe}}$\tablenotemark{a}} &
                      \colhead{Inclination} &
                     \colhead{$\Gamma$\tablenotemark{b}} &
		     \colhead{$kT_e$ (keV)} &
                     \colhead{Norm at 1 keV} & 
                     \colhead{$\Omega/2\pi$\tablenotemark{c}} &
                     \colhead{$\xi$\tablenotemark{d} }  &
                     \colhead{$\Rin$\tablenotemark{e}\ ($\Rg$)} &
\colhead{$\chi^2_\nu$} \nl
}
\startdata

1 & 30 & $1.97\pm 0.03$ & $90_{-40}^{+360}$ & $0.053$ & $0.48\pm 0.10$
  & $45_{-30}^{+45}$ & $23_{-14}^{+50}$ & 62.8/91 \nl
1 & 60 & $2.00\pm 0.03$ & $110^{+\infty}_{-57}$ & $0.055$ & $0.77^
{+0.17}_{-0.16}$
  & $13_{-8}^{+17}$ & $150^{+\infty}_{-100}$ & 68.2/91 \nl
1 & 72 & $2.01\pm 0.03$ & $140_{-80}^{+\infty}$ & $0.056$ &
  $1.20^{+0.33}_{-0.25}$ 
  & $8_{-6}^{+11}$ & $160^{+\infty}_{-100}$ & 68.4/91 \nl
2 & 30 & $1.94\pm 0.02$ & $50_{-15}^{+40}$ & $0.052$ & $0.48^{+0.10}_{-0.09}$ 
  & $10_{-9}^{+25}$ & $25^{+75}_{-14}$ & 59.6/91 \nl
2 & 60 & $1.96 \pm 0.03$ & $52^{+46}_{-17}$ & $0.053$ & 
$0.75\pm 0.16$ 
  & $0.7_{-0.7}^{+8}$ & $110_{-55}^{+\infty}$ & 61.8/91 \nl
2 & 72 & $1.97\pm 0.02$ & $55_{-20}^{+60}$ & $0.054$ & 
$1.14^{+0.21}_{-0.23}$
  & $0.01_{-0.01}^{+3}$ & $100_{-55}^{+\infty}$ & 60.3/91 \nl

\enddata
\tablenotetext{a}{Iron abundance of reflector relative to solar}
\tablenotetext{b}{Photon spectral index}
\tablenotetext{c}{Solid angle subtended by the reprocessor with respect to the
X--ray source}
\tablenotetext{d}{Ionization parameter of the reprocessor}
\tablenotetext{e}{Inner radius of the accretion disk}

\end{deluxetable}

We note that these data are consistent with previous observations of
this source (Madejski et al.\ 1995), and with the mean Seyfert spectrum
at high energies (Zdziarski et al.\ 1995). However, the 
temperatures of the Comptonizing medium 
derived from these data by Zdziarski et al. 1994) 
are rather higher ($kT\sim 250$ keV), due to their
assumption of an exponential rollover as an approximation to a
Comptonised cutoff. This also means that the derived plasma optical
depths in these previous papers of $\tau\sim 0.1$ are too low.
Modelling the spectrum with more accurate comptonised 
spectra gives $kT\sim 130$ keV and $\tau\sim 1$ (Zdziarski et al. 1996).

\section{DISCUSSION}

\subsection{Fe K Line and Overall Spectral Shape}

Firstly, we clearly see that not all AGN are consistent with a
substantial solid angle of extreme, relativistically smeared
reflection.  The reprocessed component seen in MGC--6--30--15 is not
necessarily typical of AGN in general. A similar result is seen in a
recent analysis of the ASCA spectrum of NGC 5548 (Chiang et al.\ 1999), 
where the line is broad, but not so broad as expected from a disk which 
extends down to the last stable orbit around a black hole (for the June 15th
data set they obtain $\Rin = 18.7^{+29.1}_{-9.5}\,\Rg$ for emissivity 
$\propto r^{-3}$; J.\ Chiang, private communication). 
Crucially, our data
allow us to constrain a reflected component from a molecular torus. A
torus with column $\ge 10^{23}$ cm$^{-2}$ can produce a strong, narrow
6.4 keV line, accompanied by a reflected continuum for columns $\ge
10^{24}$ cm$^{-2}$.  We do significantly detect such a contribution to
the iron K$\alpha$ line, which may also be accompanied by a reflected
continuum. However, even with a narrow line from the torus, the
remaining line from the accretion is not as broad as that seen in
MCG--6--30--15.

The amount of relativistically smeared reflection is rather less than
unity for any inclination $\le 60^\circ$. Larger inclinations are not
expected since this object is classified as a Seyfert 1 and no strong
absorption from the torus is seen in the X--ray spectrum.  Thus the
reflection from the accretion disk in IC4329a looks very like that
seen in the low state spectra of the galactic black hole systems
(\.{Z}ycki et al. 1998; Done \& \.{Z}ycki 1999) in having $\Omega/2\pi < 1$,
relativistically smeared by velocities which are inconsistent with the
reflecting material extending down to 3 Schwarzschild radii.  There is
then no intrinsic difference between the Galactic and extragalactic
accreting black holes, but there is a spread in source properties in
both classes (intrinsic spectral index, amount of reflection and
amount of relativistic smearing). The AGN results to date show that
steep spectra have a larger amount of reflection (Zdziarski et al.
1999) and more relativistic smearing.  This is exactly the sequence
seen in the Galactic Black Hole Candidate Nova Muscae 1991 (\.{Z}ycki et al.
1998; \.{Z}ycki et al. 1999) as a function of decreasing mass accretion rate.

There are currently two ways to explain the lack of extreme
relativistic line. The first is to say that the inner accretion disk
is simply not present, that it has been replaced by an X--ray hot
flow. These composite truncated disk/hot X--ray source models were
first proposed by Shapiro, Lightman and Eardley (1976) when they
discovered a hot, two temperature, optically thin solution to the
accretion flow equations, though this was subsequently shown to be
unstable.  Such models were given new impetus by the rediscovery of a
related {\it stable} solution of the accretion flow equations (Narayan
\& Yi 1995), which include advective as well as radiative cooling
(ADAFs). These ideas are clearly consistent with our results.

The alternative is that the inner disk is present, 
as required by the magnetic reconnection models for the X--ray flux,
but that it cannot be seen
in the reflected spectrum. One way to do this is if the 
upper layers of the disk are so ionized that they produce almost no atomic
spectral features (Ross et al. 1999). 
Simple models for this, where the ionisation state of the disk varies as a
smooth function of radius, do not match the data. However, the ionisation
structure could be highly complex, with rapid transition between complete 
ionisation and relatively cool material (R\'{o}\.{z}a\'{n}ska 1999; 
S. Nayakshin, private communication).
Alternatively, if the X--ray source is moving away from the disk
at transrelativistic velocities, perhaps because of plasma ejection 
from expanding magnetic loops, then its radiation pattern 
does not strongly illuminate the inner disk (Beloborodov 1999).

It is currently very difficult to distinguish observationally between
these models, and all have some remaining theoretical problems. For
the ADAF solutions, it is not yet known whether the fundamental assumptions
underlying the solutions can hold, or how a transition from the cool
disk to a hot flow can occur, while for the disk--corona geometry the
uncertainties are mainly in the detailed outcome of magnetic
reconnection, and in the ionisation structure of the illuminated disk.

\subsection{Spectral Variability}

Our data sample the source variability, which gives another way to
investigate the underlying radiation mechanisms.  This is only the second 
AGN where the reflected
and intrinsic spectrum can be disentangled (the other is NGC 5548:
Magdziarz et al.\ 1998; Chiang et al.\ 1999).  The results show
that the power law itself clearly gets intrinsically steeper as the
source brightens, which allows to constrain
the variability process. If there were merely more dissipation in
the X--ray hot corona without an accompanying change in soft seed
photon flux then the spectrum would harden as it got brighter (e.g.\
Ghisellini \& Haardt 1994). Thus
the observed spectral index--flux correlation implies that the soft
seed photons also increase, and by somewhat more than the increase in
the hard flux. Seed photons are thought to arise primarily through
reprocessing, since the hard X--rays illuminating the disk which are
not reflected are thermalised, emerging as soft photons.  In this
model the change in soft photons is commensurate with the change
in flux dissipated in the hot corona. To change the seed photons by
{\it more} than the change in hard dissipation requires either a
change in geometry, such that the hard X--ray source intercepts a
larger fraction of the disk radiation, or a decrease in reflection
albedo, so that more of the incident hard X--ray radiation is
thermalised rather than reflected. The former can be linked to the
composite hot flow/cool accretion disk models as a result of varying
the inner disk edge, while the latter could be produced in the
disk--corona models if the ionisation state of the disk
decreases for steeper spectra, so that the reflection albedo decreases
and the thermalised soft flux increases. However, recent simultaneous
observations of EUV and X--ray variability cast doubt on the simple
scenario where the EUV seed photon flux is primarily reprocessed
(Nandra et al.\ 1998; Chiang et al.\ 1999). The variability that we see
could equally well be the result of a variable soft photon flux
irradiating the X--ray region.

These three possibilities predict some differences in the behavior of
the reflected continuum. If the disk geometry is changing to give more
soft seed photons then we expect more solid angle of reflection as the
spectrum steepens (as seen in the AGN/XRB compilation of Zdziarski et
al.\ 1999).  If the increased soft photons are from increasing
thermalisation in the disk due to decreasing ionisation, then we should
also see more cold reflection (as opposed to unobservable, completely
ionized reflection) for steeper spectra. If it is simply the
irradiating soft flux which is changing, without changing disk
geometry then the reflected fraction should remain constant.

What we see is marginally (90 per cent confidence contour) consistent
with a constant reflected fraction, although the data prefer that the
relative amount of reflection {\it decreases} as the source increases
and steepens. The resulting spectrum is consistent with the absolute
normalisation of the reflected spectrum remaining constant as the
source changes. Some part of the reflected spectrum could be
contaminated by a line or reprocessed component from a molecular
torus, which would be constant due to light travel time delays on the
timescales of the monitoring campaign. Allowing for this results in
the reflected fraction from the accretion disk being more convincingly
consistent with a constant value, but still does not permit much 
of an increase with increasing flux or spectral index. The data then
support the idea of a variable soft flux which is {\it not} reprocessed
as the driver for the hard X--ray variability, but could also allow 
{\it small} changes in reflection geometry/ionisation. 

We speculate that {\it both} variability mechanisms operate in Seyferts
i.e. that there are spectral changes linked to changes in the 
geometry/ionisation (such as seen by Zdziarski et al. 1999), but that the
soft seed photons can also vary independently of these changes, giving a 
second, subtly different source of spectral variability. 

\section{CONCLUSIONS}

$\bullet$ Not all AGN have the extreme relativistic line profiles
expected from a disk extending down to the innermost stable orbit
around a black hole. 
This is consistent with either the inner disk
being truncated before the last stable orbit, or with an inner disk
which produces no significant reflected features either through
anisotropic illumination or extreme ionisation. Simple
photo--ionisation models, where the ionisation varies smoothly as a
function of radius can be ruled out by the data, but these may differ
substantially from more detailed models of the ionisation structure.

$\bullet$ There is intrinsic spectral variability, where the power law
softens as the source brightens. This implies that the soft seed
photons are increasing faster than the increase of the hard X--ray
luminosity.  The lack of a corresponding increase in the observed
reflected spectrum implies that either the changes in disk inner
radial extent/ionisation structure are small, or that the variability
is actually driven by changes in the seed photons which are decoupled
from the hard X--ray mechanism.

\section{ACKNOWLEDGEMENTS}

This research was supported in part by grant 2P03D01816 of the Polish
State Committee for Scientific Research (KBN) and by NASA grant
NAG 54106.  We thank James Chiang for discussing with us their results on
NGC 5548. We acknowledge the help of the OSSE team in reducing the
OSSE data, and the RXTE Guest Observer Facility (and in particular,
Tess Jaffe) for providing the PCA data reduction script {\tt rex}.  

\section{REFERENCES}

\ref Beloborodov A.M. 1999, ApJ, 510, L123

\ref Cappi M., Mihara T., Matsuoka M., Hayashida K., Weaver K.A., Otani C. 
1996, ApJ, 458, 149

\ref Chiang J., Reynolds C.S., Blaes O.M., Nowak M.A., Murray N.,
Madejski G.M., Marshall H.L., Magdziarz P. 1999, ApJ, in press

\ref di Matteo T. 1998, MNRAS, 299, 15

\ref Done C., \.{Z}ycki P.T. 1999, MNRAS, 305, 457

\ref Esin A. A., McClintock J.E., Narayan R. 1997, ApJ, 489, 865

\ref Fabian A.C., Rees M.J., Stella L., White, N.E. 1989, MNRAS, 238, 729

\ref Galeev A.A., Rosner R., Vaiana G.S. 1979, ApJ, 229, 318

\ref George I.M., Fabian A.C. 1991, MNRAS, 249, 352

\ref George I.M., Turner T.J., Netzer H., Nandra K., Mushotzky R.F., Yaqoob T.
ApJS, 114, 73

\ref Ghisellini G., Haardt F. 1994, ApJ, 429 L53

\ref Ghisellini G., Haardt F., Matt G. 1994, MNRAS, 267, 743

\ref Gierli\'{n}ski M., Zdziarski A. A., Done C., Johnson W. N.,  Ebisawa
K., Ueda Y., Phlips F. 1997, MNRAS, 288, 958

\ref Gierli\'{n}ski M., Zdziarski A. A., Poutanen J., Coppi P. S., 
Ebisawa K., Johnson W. N. 1999, MNRAS, in press

\ref Haardt F., Maraschi L., Ghisellini G. 1994, ApJ, 432, 95

\ref Henry R.B.C., Worthey G. 1999, PASP, 111, 919

\ref Iwasawa K., Fabian A.C., Mushotzky R.F., Brandt W.N., Awaki H.,
Kunieda H. 1996, MNRAS, 279, 837

\ref Janiuk A., \.{Z}ycki P.T., Czerny B. 1999, MNRAS, submitted

\ref
Johnson W.N., Zdziarski A.A., Madejski G.M., Paciesas W.S., Steinle H.,
Lin Y.-C., 1997, in Dermer C. D., Strickman M. S., Kurfess J. D., eds, 
The 4th Compton Symposium, AIP, New York, 283

\ref Krolik J.H., Madau P., \.{Z}ycki P.T. 1994, ApJ, 420, L57

\ref Lightman A.P., White T.R. 1988, ApJ, 335, 57

\ref Lightman A.P., Zdziarski A.A. 1987, ApJ, 319, 643

\ref Madejski, G.M.,   Sikora, M., Jaffe, T., B{\l}a\.{z}ejowski, M., Jahoda,
K., Moderski, R. 1999, ApJ, in press

\ref Madejski, G. M., et al. 1995, ApJ, 438, 672

\ref Magdziarz P., Zdziarski A.A., 1995, MNRAS, 273, 837

\ref Magdziarz P., Blaes O.M., Zdziarski A.A., Johnson W.N., Smith D.A.
1998, MNRAS, 301, 179

\ref Makishima K., et al., 1996, PASJ, 48, 171

\ref Matt G., Perola G.C.,  Piro L. 1991, A\&A, 247, 25

\ref Matt G., Fabian A.C., Ross R., 1993, MNRAS, 262, 179

\ref Morrison R., McCammon D. 1983, ApJ, 270, 119

\ref Mushotzky R.F., Fabian A.C., Iwasawa K., Kunieda H., Matsuoka M., 
Nandra K., Tanaka Y. 1995, MNRAS, 272, 9p

\ref Nandra K., Pounds K.A. 1994, MNRAS, 268, 405

\ref Nandra K., George I.M., Mushotzky R.F., Turner T.J., Yaqoob T. 1997,
ApJ, 477, 602

\ref Nandra K., Clavel J., Edelson R.A., George I.M., Malkan M.A., 
Mushotzky R.F., Peterson B.M., Turner T.J. 1998, ApJ, 505, 594

\ref Narayan R., Yi I. 1995, ApJ, 444, 231

\ref Pounds K.A., Nandra K., Stewart G.C., George I.M., Fabian A.C. 1990,
Nature, 344, 132

\ref Press W.H., Teukolsky S.A., Vettering W.T., Flannery B.P. 1992,
Numerical Recipes (New York: Cambridge University Press)

\ref Reynolds C.S. 1997, MNRAS 286, 513

\ref Ross R.R., Fabian A.C., Young A.J. 1999, MNRAS, 306, 461

\ref Rothschild, R.E., et al. 1998, ApJ, 496, 538

\ref R\'{o}\.{z}a\'{n}ska A. 1999, MNRAS, 308, 751

\ref Rybicki G.B., Lightman A.P. 1979, Radiative Processes in Astrophysics
(New York: Wiley-Interscience)

\ref Shakura, N.I., Sunyaev, R.A. 1973, A\& A, 24, 337

\ref Shapiro S.L., Lightman A.P., Eardley D.M. 1976, ApJ, 204, 187

\ref Tanaka Y. et al. 1995, Nature, 375, 659

\ref Toor A., Seward F.D., 1974, AJ., 79, 995

\ref Weaver K.A., Yaqoob T., Mushotzky R.F., Nousek J., Hayashi I., Koyama K. 
1997, ApJ, 474, 675

\ref Weaver K.A., Reynolds C. 1998, ApJ, 508, L39

\ref Zdziarski A.A., Fabian A.C., Nandra K., Celotti A., Rees M.J., Done C., 
Coppi P., Madejski G.M. 1994, MNRAS, 269, 55p

\ref Zdziarski A.A., Johnson W.N., Done C., Smith D.A., McNaron--Brown K.
1995, ApJ, 438, 63

\ref Zdziarski A.A., Gierli\'{n}ski M., Gondek D., Magdziarz P. 1996,
A\&AS, 
120, 553

\ref Zdziarski A.A., Lubi\'{n}ski P., Smith D.A. 1999, MNRAS, 303, L11

\ref Zhang, W., Giles, A.B., Jahoda, K., Swank, J. H, Morgan, E.M. 1993,
in ``EUV, X--ray, and Gamma-ray Instrumentation for Astronomy IV,'' SPIE
Proceedings, O. Siegmund ed., 2006, 324

\ref \.{Z}ycki P.T., Done C., Smith D.A. 1997, ApJ, 488, L113

\ref \.{Z}ycki P.T., Done C., Smith D.A. 1998, ApJ, 496, L25

\ref \.{Z}ycki P.T., Done C., Smith D.A. 1999, MNRAS, 305, 231

\end{document}